\documentclass[pra,twocolumn,nofootinbib,floatfix,10pt]{revtex4-2}

\usepackage{amsmath}
\usepackage{amssymb}
\usepackage{wasysym}
\usepackage{graphicx}
\usepackage{color,soul}
\usepackage{physics}
\usepackage{siunitx}
\usepackage{dsfont}
\usepackage{float}
\usepackage[english]{babel}
\usepackage{blindtext}
\usepackage[english,nomargin,inline,marginclue,draft]{fixme}
\pdfpageheight\paperheight
\pdfpagewidth\paperwidth

\usepackage[colorlinks,linkcolor=blue,anchorcolor=blue,citecolor=blue,urlcolor=blue]{hyperref}

\fxusetheme{colorsig}
\FXRegisterAuthor{cg}{acg}{CG}  
\FXRegisterAuthor{th}{ath}{\color{blue}TH}  
\FXRegisterAuthor{ib}{aib}{\color{red}IB} 
\FXRegisterAuthor{sh}{ash}{\color{cyan}SH} 
\FXRegisterAuthor{db}{adb}{\color{green}DB} 
\FXRegisterAuthor{ps}{aps}{PS}
\makeatletter
\renewcommand*\FXLayoutInline[3]{%
  {\@fxuseface{inline}\ignorespaces{\color{fx#1}[#3: #2]}}}
\makeatother

\long\def\symbolfootnote[#1]#2{\begingroup%
\def\thefootnote{\fnsymbol{footnote}}\footnotetext[#1]{#2}\endgroup}

\def\nobreakbefore{%
  \relax\ifvmode\else
    \ifhmode
      \ifdim\lastskip > 0pt\relax
        \unskip\nobreakspace
      \else 
        \nobreakspace
      \fi
    \fi
  \fi
}
\let\oldcite\cite
\renewcommand\cite{\nobreakbefore\oldcite}

\begin{document}
\title{Bifurcation of time crystals in driven and dissipative Rydberg atomic gas}

\author{Bang Liu$^{1,2}$}
\author{Li-Hua Zhang$^{1,2}$}
\author{Zong-Kai Liu$^{1,2}$}
\author{Jun Zhang$^{1,2}$}
\author{Zheng-Yuan Zhang$^{1,2}$}
\author{Shi-Yao Shao$^{1,2}$}
\author{Qing Li$^{1,2}$}
\author{Han-Chao Chen$^{1,2}$}
\author{Yu Ma$^{1,2}$}
\author{Tian-Yu Han$^{1,2}$}
\author{Qi-Feng Wang$^{1,2}$}
\author{Dong-Sheng Ding$^{1,2,\textcolor{blue}{\dag}}$}
\author{Bao-Sen Shi$^{1,2}$}

\affiliation{$^1$Key Laboratory of Quantum Information, University of Science and Technology of China; Hefei, Anhui 230026, China.}
\affiliation{$^2$Synergetic Innovation Center of Quantum Information and Quantum Physics, University of Science and Technology of China; Hefei, Anhui 230026, China.}

\date{\today}

\symbolfootnote[2]{dds@ustc.edu.cn}

\maketitle

\textbf{A time crystal is an exotic phase of matter where time-translational symmetry is broken; this phase differs from the spatial symmetry breaking induced in crystals in space. Lots of experiments report the transition from a thermal equilibrium phase to time crystal phase. However, there is no experimental method to probe the bifurcation effect of distinct continuous time crystals in quantum many-body systems. Here, in a driven and dissipative many-body Rydberg atom system, we observe multiple continuous dissipative time crystals and emergence of more complex temporal symmetries beyond the single time crystal phase. Bifurcation of time crystals in strongly interacting Rydberg atoms is observed; the process manifests as a transition from a time crystal state of long temporal order to one of short temporal order, or vice versa. By manipulating the driving field parameters, we observe the time crystal’s bistability and a hysteresis loop. These investigations indicate new possibilities for control and manipulation of the temporal symmetries of non-equilibrium systems. }

The term time crystal refers to a unique state of matter in which a system exhibits spontaneous breaking of its time translation symmetry; this state was initially proposed by Frank Wilczek \cite{wilczek2012quantum}. In other words, the time crystal is a phase of matter in which the system's behavior repeats periodically in both space and time. A time crystal exhibits long-range order in time, whereas conventional crystals exhibit long-range order in space. The state includes discrete and continuous time crystal phases \cite{sacha2017time,else2020discrete,kongkhambut2022observation,zaletel2023colloquium} that exhibit discrete and continuous time-translation symmetry breaking, respectively. Since the initial proposal, time crystals were first reported in experiments using trapped ions \cite{zhang2017observation} and the nitrogen-vacancy centers in diamonds \cite{choi2017observation}. There has been considerable progress on both the theoretical \cite{watanabe2015absence,syrwid2017time,huang2018clean,gong2018discrete,yao2020classical,time-crystalline-transitions,gambetta2019discrete} and experimental fronts in efforts to study and understand time crystals \cite{li2012space,else2016floquet,autti2018observation,smits2018observation,pizzi2021bistability,autti2021ac,trager2021real}. Researchers continue to strive to improve the stability \cite{machado2023absolutely} and the coherence \cite{rovny2018observation,randall2021many} of time crystals, and to probe the equilibrium of quantum matter \cite{ho2017critical}. Interestingly, the system comes out of thermal equilibrium dissipation under external driving conditions, and researchers have observed stabilized dissipative time crystals \cite{kessler2021observation}, prethermal discrete time crystals \cite{vu2023dissipative,kyprianidis2021observation}, and higher-order time crystals \cite {pizzi2019period,pizzi2021higher}.

The large dipole moment of the Rydberg atom provides a good platform for study of the dynamics of strongly correlated systems. For example, researchers have used Rydberg atoms to probe quantum phase transitions \cite{bernien2017probing,keesling2019quantum}, quantum scars \cite{serbyn2021quantum,bluvstein_controlling_2021}, non-equilibrium phase transitions, and self-organized criticality \cite{lee2012collective,carr2013nonequilibrium,ding2019Phase,helmrich2020signatures,klocke2021hydrodynamic,ding2022enhanced}. In traditional equilibrium systems, the second law of thermodynamics dictates that entropy will always increase with time, and this leads to thermal equilibrium and the absence of persistent oscillatory behavior. In the driven and dissipative Rydberg system, however, the external driving force injects energy into the system, while the long-range interactions between the system's constituents aid in redistributing this energy and thus help to maintain a persistent oscillatory evolution of the Rydberg atom population \cite{ding2023ergodicity, wadenpfuhl2023emergence}, and generating dissipative continuous time crystal \cite{wu2023observation}. The driven and dissipative Rydberg atom system is inherently an out-of-equilibrium complex system, which means that it does not settle into a steady state or thermal equilibrium, and the exotic emergence and transition dynamics of distinct continuous time crystals are worthy of study. 

In this work, we observe a bifurcation effect for the continuous time crystals in strongly interacting Rydberg atoms under external radio-frequency (RF)-field continuous driving. The bifurcation of time crystals in Rydberg atoms originates from the competition between the different Floquet resonances. Coupling between the (multiple) many-body states leads to the formation of distinct time crystals, which subsequently results in the emergence of more complex temporal symmetries beyond the simple discrete-time translation symmetry. We observe the full phase diagram of the system, including multiple distinct time crystal phases with several periodicities and chaotic phases. In addition, we observe phase transitions between the distinct time crystals and the hysteresis loop. Study of the emergence of these distinct time crystals will provide new possibilities for exploration of the dynamics and control of quantum many-body systems when out of equilibrium. 

\begin{figure*}
\centering
\includegraphics[width=2.08\columnwidth]{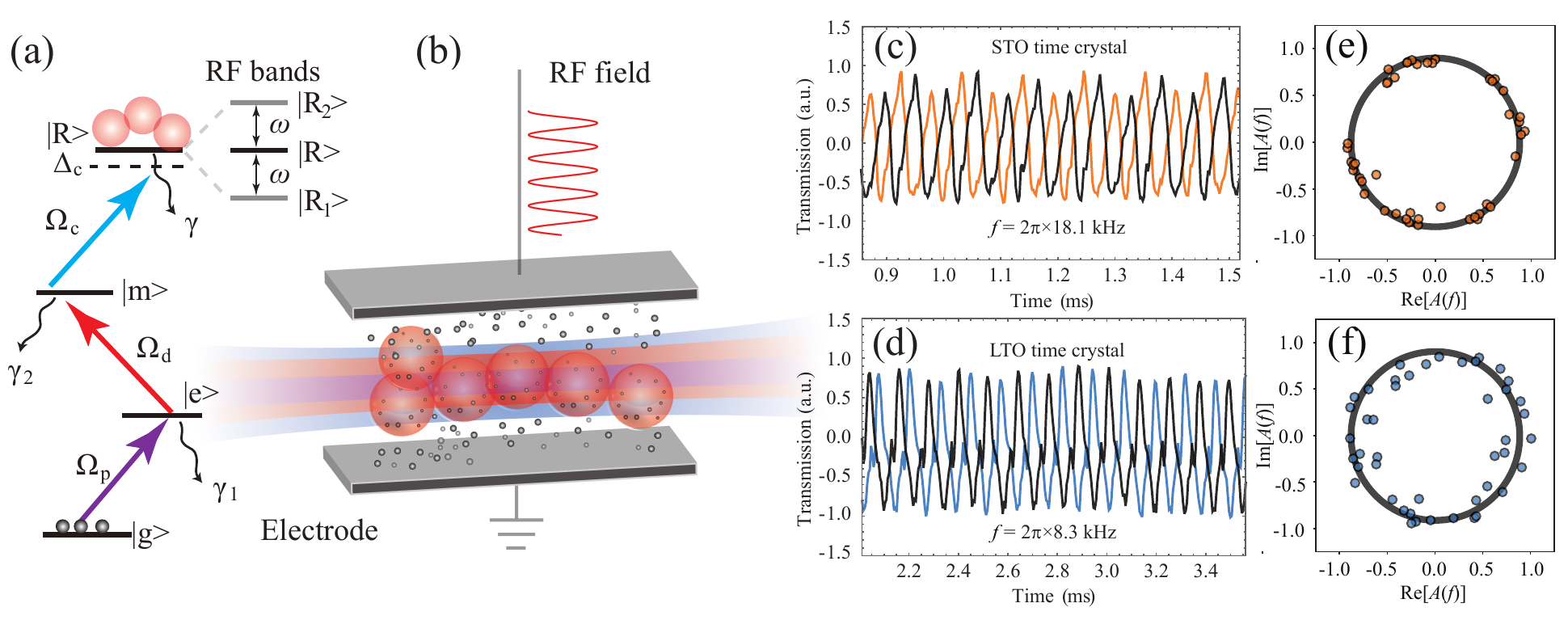}\\
\caption{\textbf{Energy level diagram and experimental setup for time crystal measurement.} (a) Energy level diagram based on three-photon Rydberg electromagnetically-induced transparency (EIT) scheme. A probe laser with Rabi frequency $\Omega_{p}$ drives the transition from the cesium atom ground state $\ket{g}$ to intermediate state $\ket{e}$ ($\ket{6S_{1/2}}$ to $\ket{6P_{3/2}}$). The dressing laser with Rabi frequency $\Omega_{d}$ drives the transition from intermediate state $\ket{e}$ to another intermediate state, $\ket{m}$ ($\ket{6P_{3/2}}$ to $\ket{7S_{1/2}}$). The coupling laser with Rabi frequency $\Omega_{c}$ and detuning $\Delta_c$ drives the transition from the intermediate state $\ket{m}$ to the Rydberg state $\ket{R}$ ($\ket{7S_{1/2}}$ to $\ket{49P_{3/2}}$). $\gamma_1$, $\gamma_{2}$, and $\gamma$ correspond to the decay rates of states $\ket{e}$, $\ket{m}$, and $\ket{R}$, respectively. The Rydberg state $\ket{R}$ is divided into Floquet sidebands when atoms are driven by an RF field; three sideband energy levels, $\ket{R}$, $\ket{R_1}$, and $\ket{R_2}$ with energy interval $\omega$, are illustrated. (b) Simplified experimental setup. An RF field is applied to the atoms by two electrodes with a loading frequency ranging from 0 MHz $\sim$ $2\pi\times$30 MHz. (c), (d) Triggered probe transmission caused by switching on the RF field [$U$ = 3.8 V], which oscillates with distinct frequencies [$f = 2\pi\times18.1$ kHz (c) and $f = 2\pi\times8.3$ kHz (d)] obtained by setting $\Delta_c = -2\pi\times 24.1$ MHz for (c) and $\Delta_c = -2\pi\times 32.4$ MHz for (d), thus revealing the distinct time crystals. The different colored lines in (c) and (d) are from different experimental trials. (e), (f) Distributions of the Fourier amplitudes with the dominant frequency on the complex plane. In this process, we recorded the probe transmission within the time intervals $\Delta t=0.66$ ms (e) and $\Delta t=1.55$ ms (f) with 300 data points after the RF-field is turned on for (e) 0.858 ms and (f) 2.01 ms.}
\label{setup}
\end{figure*}

\subsection*{Physical model and experiment setup}
To determine the time translation symmetry breaking mechanism in driven and dissipative Rydberg atoms, we construct a simplified physical model to simulate the observations that considers $N$ interacting two-level atoms with a ground state $\ket{g}$ and the Rydberg state $\ket{R}$ (with a decay rate $\gamma$). A laser couples the atoms with a Rabi frequency $\Omega$ and detuning $\Delta$. The Rydberg atoms are affected by the many-body interaction strength $V = {C_6}/{r^6} $ [where $C_6$ is the van der Waals coefficient and $r$ represents the distance between the Rydberg atoms]. These atoms are exposed to an external RF field with an electric field component $E_{\rm{RF}}$ and frequency $\omega$. The RF field perturbs the system and induces additional frequencies into the system spectrum, thus leading to the appearance of RF sidebands of the Rydberg states \cite{miller2016radio}; see the Methods section for further details. In our model, we consider the $\pm 1$-order sidebands of the Rydberg states as $\ket{R_{1,2}}$ with the corresponding Rabi frequency $\Omega_{1,2}$ and detuning $\Delta_{1,2}$. We use a mean field treatment to simulate multiple periodic oscillations in the Rydberg atom population; see the Methods section for further details.

In the experiments, we excite cesium atoms in a thermal vapor to study the features of time crystals. The cesium atom energy level structure and the experimental setup are depicted in Figs.~\ref{setup}(a) and (b), respectively. We use a three-photon electromagnetically-induced transparency (EIT) scheme to prepare the Rydberg atoms, as described in a previous study \cite{zhangRydberg,liuHighly}. Specifically, the excitation process involves use of a 852 nm probe beam to resonantly drive the transition from state $\ket{6S_{1/2}}$ to state $\ket{6P_{3/2}}$ with a Rabi frequency $\Omega_p$, a resonant 1470 nm laser with a Rabi frequency $\Omega_d$ to drive the transition from state $\ket{6P_{3/2}}$ to state $\ket{7S_{1/2}}$), and a 780 nm coupling beam with a Rabi frequency $\Omega_c$ and detuning $\Delta_c$ to drive the transition from state $\ket{7S_{1/2}}$ to $\ket{49P_{3/2}}$. The Rydberg atoms are illuminated by the RF field (the AC stark effect for the atomic state with the low principal quantum number can be ignored in this case). By tuning the system's parameters, we can observe the distinct continuous time crystals, as shown in Fig.\ref{setup}(c) and Fig.\ref{setup}(d). 

\begin{figure*}
\centering
\includegraphics[width=2\columnwidth]{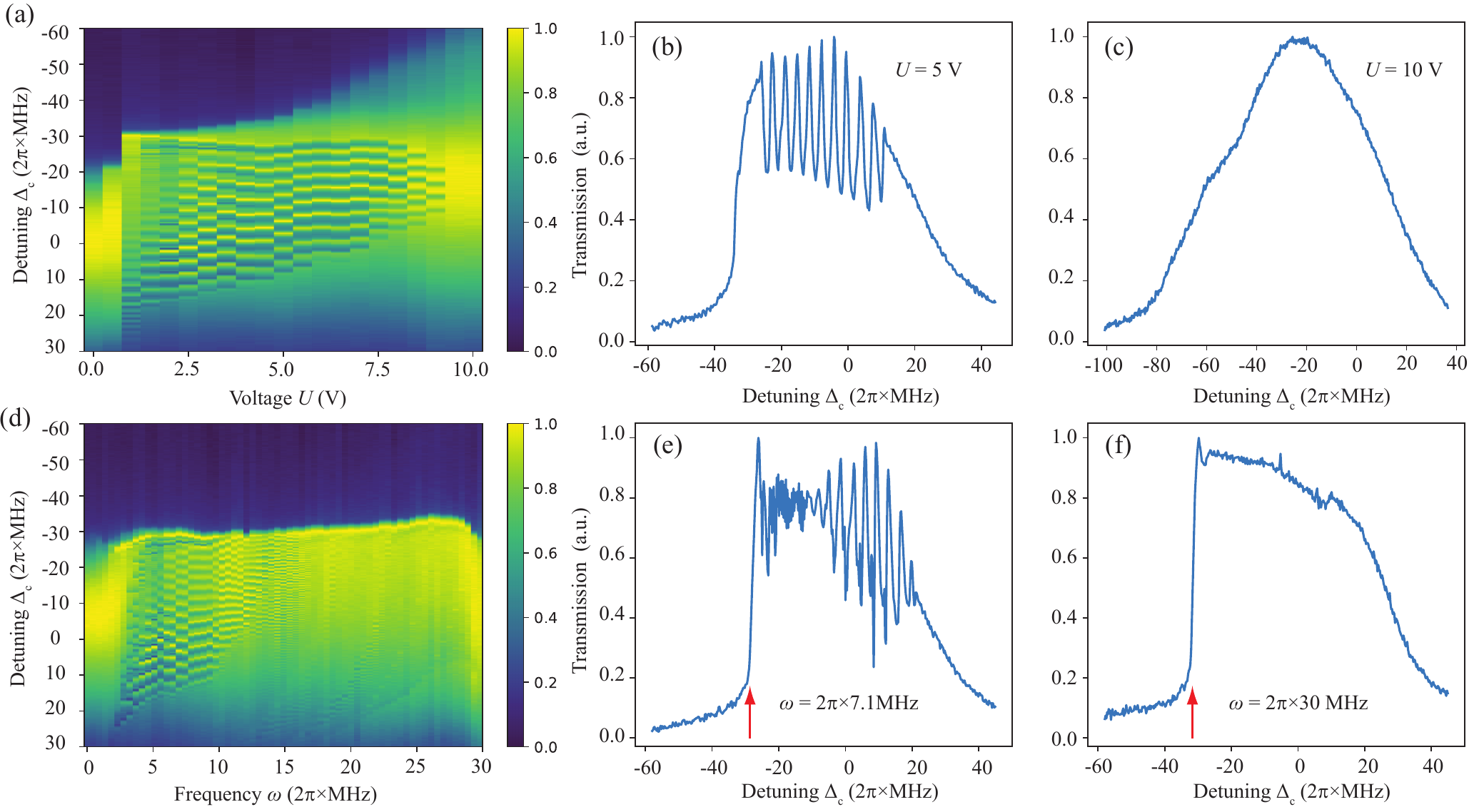}\\
\caption{\textbf{Measured phase diagram.} Color maps of probe transmission (normalized with respect to their respective maximal values) when scanning $\Delta_c$ and the external RF field voltage $U$ (a), and the external RF field frequency $\omega$ (d). Here, the coupling detuning $\Delta_c$ is scanned from $\Delta_c=-2\pi\times $60 MHz to $\Delta_c=2\pi\times $30 MHz. The voltage is scanned from $U$ = 0 to $U$ = 10 V (a) and the frequency changes from $\omega = 0$ to $\omega = 2\pi\times$ 30 MHz (d). In the oscillating regions in panels (a) and (d), the system is in the time translation symmetry breaking phase. The color bar ranging from 0 to 1 represents the transmission intensity. (b) and (c) correspond to probe transmissions with $U$ = 5 V and $U$ = 10 V, respectively. Here, the RF field is set at $\omega= 2\pi\times7.2$ MHz. (e) and (f) are the probe transmissions with RF field frequencies of $\omega= 2\pi\times7$ MHz and $\omega= 2\pi\times28$ MHz, respectively. The voltage is set $U$ = 2 V. The red arrows shown in (e) and (f) indicate sudden jumps in the probe transmission spectrum, which are called non-equilibrium phase transitions. In panels (b) and (e), oscillation behavior occurs in the probe transmission spectrum that corresponds to the non-trivial regime for time translation symmetry breaking.}
\label{phase diagram}
\end{figure*}

Figure \ref{setup}(c) represents the short temporal order (STO) time crystal and Fig.\ref{setup}(d) represents the long temporal order (LTO) time crystal. In addition, we present evidence of a random phase distribution in these distinct time crystals, which varies naturally from aperiodic to periodic. By measuring the dominant frequency peak $f$ in the Fourier spectrum from 45 repeated trials (Fig.\ref{setup}(e)) and (Fig.\ref{setup}(f)) [see further details in the Methods section], we obtain the distribution of the Fourier amplitudes in the complex plane. The phase of the Fourier amplitude $A(f)$ at the dominant frequency $f$ is distributed randomly between 0 and 2$\pi$, as shown in Fig.\ref{setup}(e) and Fig.\ref{setup}(f), thus indicating spontaneous breaking of the continuous time translation symmetry \cite{kongkhambut2022observation}.

\subsection*{Phase diagram}
To probe the exotic phase in the Rydberg atoms under RF field driving conditions, we measured the system’s phase diagram. The phase diagram provides a comprehensive map of the different phases or states in which the transmission can exist as a function of various parameters, e.g., the coupling detuning $\Delta_c$, voltage $U$, and frequency $\omega$ of the RF field. We varied the RF field intensity by varying the voltages of the electrodes and measured the probe transmission versus the coupling detuning $\Delta_c$. Figure \ref{phase diagram}(a) shows the color map of the probe transmission when $U$ = 0 $\sim$ 10 V and $\Delta_c$= -$2\pi\times $60 MHz $\sim$ $2\pi\times $30 MHz. Figure \ref{phase diagram}(b) and (c) show examples of the probe transmission at $U$ = 5 V and $U$ = 10 V, respectively. We also changed the frequency of the RF field from $\omega= 0$ MHz to $\omega= 2\pi\times30$ MHz and recorded the probe transmission, with results as shown in Fig. \ref{phase diagram}(d). Figure \ref{phase diagram}(e) and (f) show the corresponding results with $\omega= 2\pi\times7$ MHz and $\omega= 2\pi\times28$ MHz, respectively.

\begin{figure*}
\centering
\includegraphics[width=2.08\columnwidth]{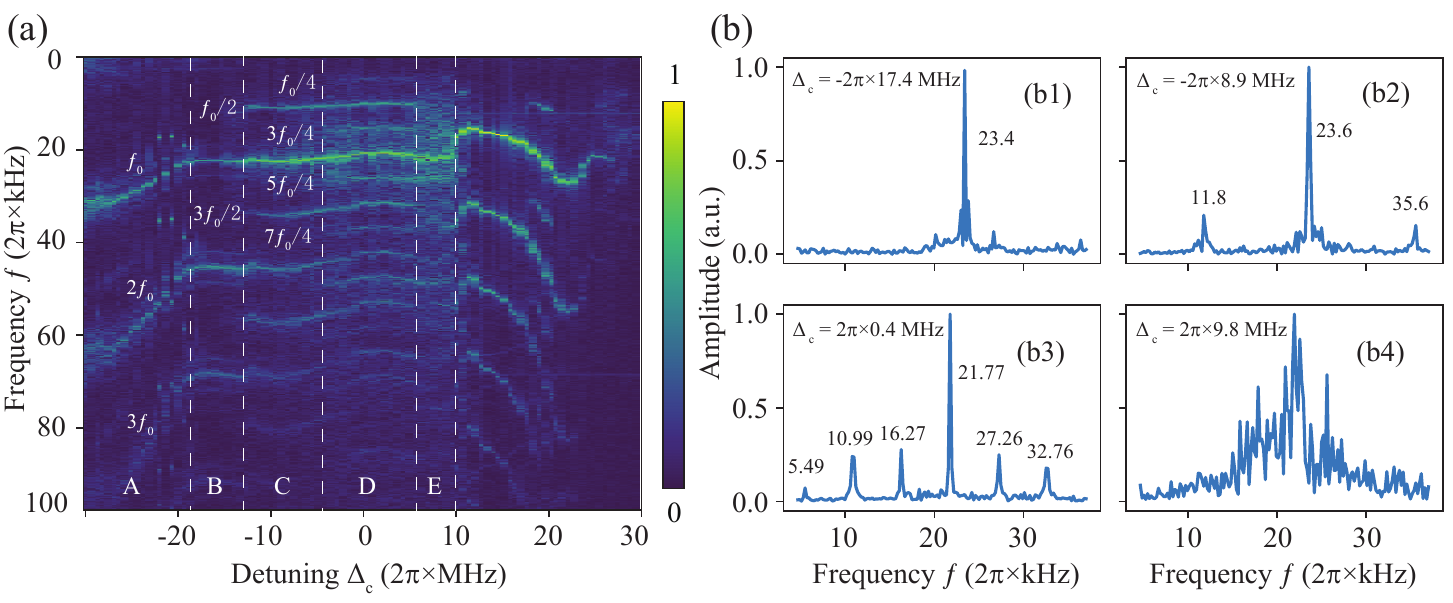}\\
\caption{\textbf{Distinct time crystals.} (a) Measured Fourier spectrum versus the coupling detuning $\Delta_{c}$. Labels A-E in (a) indicate the areas that characterize the distinct time crystal phases. The measured responses of several non-integer multiples of $f = f_0$ represent the `high-order'  continuous time crystals relative to the $f = f_0$ time crystal. The color bar from 0 to 1 represents the Fourier transform intensity. (b) is the corresponding Fourier spectrum at the detuning $\Delta_{c}/2\pi$ = -17.4 MHz (b1), $\Delta_{c}/2\pi$ = -8.9 MHz (b2), $\Delta_{c}/2\pi$ = -1.3 MHz (b3), and $\Delta_{c}/2\pi$ = 9.8 MHz (b4), respectively. The peaks in Fourier spectra (b1)-(b3) show the evidence of distinct periodicity, and (b4) is the measured Fourier spectrum of chaotic phase. }
\label{Distinct time crystal phases}
\end{figure*}

From these results, we found that oscillation effects occurred within the $U$ = 1 V $\sim$ $U$ = 8.5 V range [Fig. \ref{phase diagram}(a)] and the $\omega= 2\pi\times2$ MHz $\sim$ $\omega= 2\pi\times16$ MHz range [Fig. \ref{phase diagram}(d)]. For example, in Fig. \ref{phase diagram}(b), the system response exhibits repetitive back-and-forth motion on probe transmission at around resonance, whereas no oscillation is shown in Fig. \ref{phase diagram}(c). In Fig. \ref{phase diagram}(a), the area of the oscillation covers a large coupling detuning when $U$ = 1 V, but becomes narrower with increasing voltage $U$. Above $U$ = 8.5 V, the probe transmission becomes normal. The oscillation’s disappearance at high $U$ and high frequency $\omega$ can be attributed to: (i) the high-intensity RF field inducing broadening larger than the shift from the Rydberg atom interactions; and (ii) the RF-field-induced Floquet energy interval (which is proportional to $\omega$) being larger than the shift from the Rydberg atom interactions at high $\omega$. In addition, there are sudden jumps indicated by red arrows in Fig. \ref{phase diagram}(e) and Fig. \ref{phase diagram}(f). These jumps are non-equilibrium phase transitions from non-interacting Rydberg atoms to interacting Rydberg atoms \cite{ding2019Phase}, in which the broadening effect induced by the RF field enables facilitated excitation of the Rydberg atoms. 

\subsection*{Distinct time crystal phases}
To study the characteristics of the time crystal phases in our system, we measured the Fourier spectrum of the oscillated transmission, as shown in Fig. \ref{Distinct time crystal phases}(a). There are comb-shaped spectra and chaotic spectra distributed in the varied $\Delta_c$, which are separated within areas A-E. In area A, we see that the first peak ($f=f_0$) of the Fourier spectrum is nearly linear versus the coupling detuning $\Delta_c$. In area B, the frequency of the time crystal remains stable versus varying $\Delta_c$; see the peak in panel (b1) of Fig. \ref{Distinct time crystal phases}(b). Because the time crystal generally exists in the low energy or many-body ground states, the change in parameters does not alter the many-body state of the Rydberg atoms in area B, thus resulting in a long-term stable time crystal. Interestingly, the time crystal becomes bifurcated when $\Delta_c$ is increased further. For example, a non-integer multiple of the $f_0$ frequency signals appears in areas C and D. In panels (b2) and (b3), there are a series of branches at $f=f_0/4, f_0/2, 3f_0/4, 5f_0/4, 3f_0/2$, and $7f_0/4$ at around $f = f_0$ in the Fourier spectrum. 

\begin{figure*}
\centering
\includegraphics[width=2.08\columnwidth]{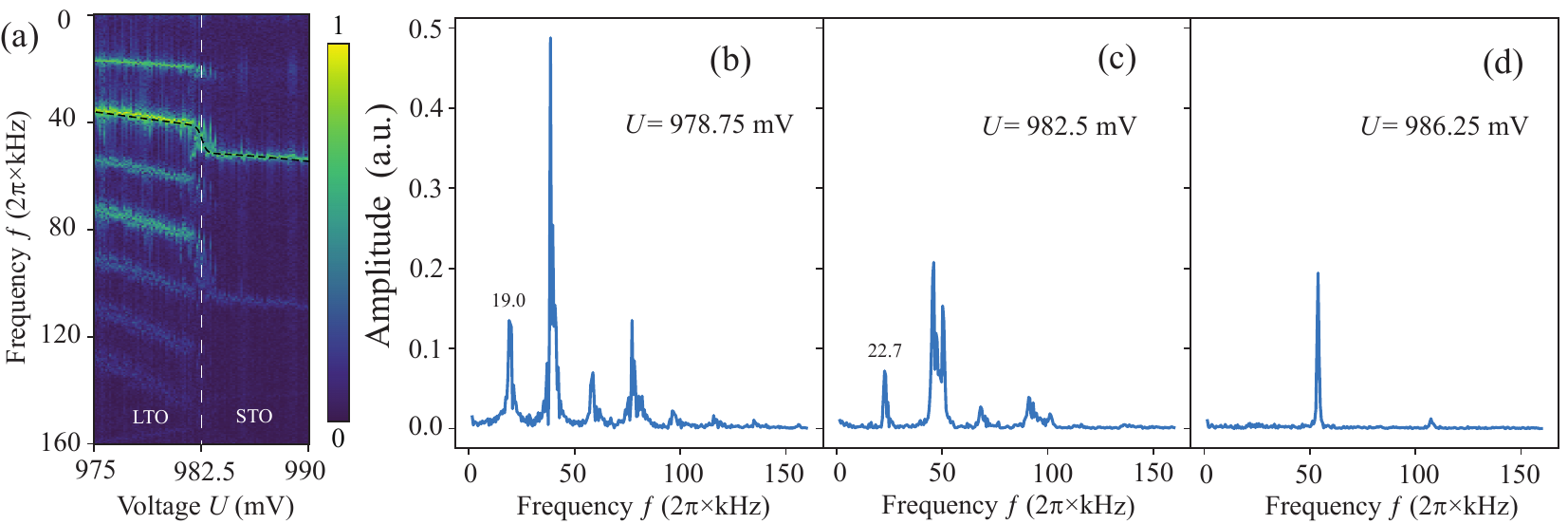}\\
\caption{\textbf{Phase transition between distinct time crystals.} (a) Phase diagram of a voltage range from $U$ = 975 mV to $U$ = 990 mV in a high resolution. The black dashed line represents the fitting function $f=x_0+kU(d(e^{b+aU}$-1))/(
 1+$e^{b+aU}(a_1U+b_1$)) with $x_0$=-659.9, $d$=5.8, $k$=720, $a$=5333.33, $b$=-5240, $a_1$=160, and $b_1$=-156. The color bar from 0 to 1 represents the Fourier transform intensity. (b)-(d) Corresponding Fourier spectra at $U$ = 978.75 mV (b), $U$ = 982.5 mV (c), and $U$ = 986.25 mV (d). The phase diagram includes two temporal orders called the LTO and the STO, which are separated by the white dashed line. In the STO time-crystal phase, the temporal order is the shorter-range order, which means that the system's response is repeated periodically over shorter time intervals. In the LTO time-crystal phase, however, the system's response is repeated periodically over longer periods of time.}
\label{phase transition}
\end{figure*}

The emergence of the non-integer multiple of $f=f_0$ frequency signals means that the Rydberg atom population oscillates between more than two energy states with multiple specific periods, thus creating new temporal orders. Increasing the coupling detuning from $\Delta_c = -2\pi\times30$ MHz to $\Delta_c = -2\pi\times0$ MHz adds to the population of Rydberg atoms and thus increases the interactions between the Rydberg atoms. This change in the interaction enables generation of a low energy gap in the temporal order, leading to formation of a time crystal state with an LTO.

By increasing $\Delta_c$ even further [e.g., to $\Delta_c = 2\pi\times 6$ MHz], the system enters a chaotic regime in which the time crystals bifurcate further; see the relatively chaotic frequency spectrum in area E. Panel (b4) in Fig. \ref{Distinct time crystal phases}(b) shows an example of the Fourier spectrum in area E. The bifurcation leads to formation of additional temporal order patterns and breakdown of the existing temporal order. This causes the excitation of the Rydberg atoms to be far from synchronized. As a result, the system may show more complex and unpredictable dynamics, along with the emergence of new frequency components and irregular oscillations. When $\Delta_c > 2\pi\times 10$ MHz, the system regains the single trajectory for the time crystal phase as the interactions become weak.

\subsection*{Phase transition of time crystal}
We also measured the response versus the RF field intensity and mapped the phase diagram of the voltage range from 975 mV to 990 mV [see Fig.~\ref{phase transition}(a)], and this diagram includes two stable time crystals with different periodicities and their transition. When the voltage is in the 975 mV $<$ $U$ $<$ 982.5 mV range, the system has a low-frequency time crystal phase. For example, we measured the Fourier spectrum at $U$ = 978.75 mV; there was a low-frequency peak at $f=2\pi\times19.0$ kHz, as shown in Fig.~\ref{phase transition}(b). With increasing voltage $U$, we found that the onset of the high-frequency time crystal phase appears on exceeding the critical voltage $U$ = 982.5 mV. The low-frequency component in the Fourier spectrum then disappears and the high-frequency term remains when we increase from $U$ = 975 mV to $U$ = 990 mV; these results are shown in Fig.~\ref{phase transition}(b)-(d).

Varying the RF field intensity causes the system's symmetry to vary and the probe transmission then shows distinct periodicity. The phase transition represents the system changing from an LTO time-crystal into an STO time-crystal. We characterized the scaling of this phase transition using a fitting function; see the dashed line and the description in the caption of Fig.~\ref{phase transition}(a). This function supports a continuous transition from an LTO time crystal to an STO time crystal.

\subsection*{Bistability of the time crystal}
\begin{figure*}
\centering
\includegraphics[width=2.08\columnwidth]{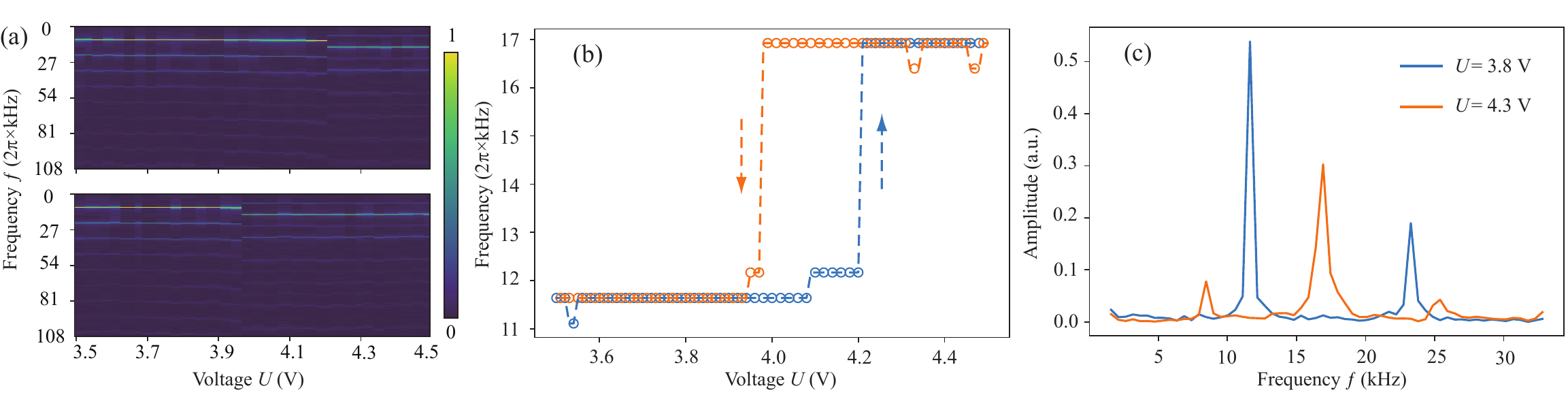}\\
\caption{\textbf{Bistability of time crystal}. (a) Measured phase diagrams with forward scanning from $U$ = 3.5 V to $U$ = 4.5 V and vice versa. (b) Measured frequency of the dominant peak versus voltage $U$ from (a). The blue data represent the forward scanning case and the orange data correspond to the backward scanning case. There is a hysteresis loop that indicates the relationship between the voltage $U$ applied to the Rydberg atoms and the resulting dominant peak in the Fourier spectrum of the probe transmission. (c) Fourier spectra for $U$ = 3.8 V (blue) and $U$ = 4.3 V (orange).}
\label{Bistability}
\end{figure*}

By scanning the RF field intensity in the forward and backward directions, we observed the bistability of the time crystal; the results are shown in Fig.~\ref{Bistability}(a). Here, in the context of time crystals, bistability refers to the existence of two stable time crystal states or many-body energy levels that the Rydberg system can switch between. The upper panel in Fig.~\ref{Bistability}(a) corresponds to the measured phase diagram when scanning the voltage $U$ positively, and the lower panel in Fig.~\ref{Bistability}(a) shows the phase diagram measured when scanning the voltage $U$ negatively. When $U$ is increased gradually, the time crystal phase is stable at a fixed frequency $f \sim 2\pi\times11.64$ kHz within the range from $U$ = 3.5 V to $U$ = 4.2 V. Then, the system's response shows a sudden jump when it crosses the critical point at $U$ = 4.2 V, and the system is then stable in another time crystal phase with a stable frequency of $f \sim 2\pi\times16.93$ kHz; see the results shown in Fig.~\ref{Bistability}(b). Similarly, when $U$ is gradually reduced, the time crystal phase switches at another entirely different critical point at $U$ = 3.95 V, as shown by the orange data in Fig.~\ref{Bistability}(b). 

The measured critical point voltages $U$ during forward and backward scanning differ because the time crystal has a memory effect that indicates its ability to retain specific oscillatory behavior (i.e., preserving the memory of its original state). This results in a delayed response in the RF field. This response lag leads to hysteresis and loop formation, as shown in Fig.~\ref{Bistability}(b). Figure.~\ref{Bistability}(c) illustrates the measured Fourier spectra of these two distinct time crystals.

\subsection*{Discussion}

The observed time crystals are characterized using their nontrivial temporal orders, which represent the non-equilibrium persistent oscillations of the excited Rydberg atom population. There are few previous experimental reports of observation of the bifurcation effect of time crystals. The multiple time crystals and the bistability observed here open up avenues to study non-equilibrium physics with dependence on distinct temporal orders. Furthermore, the existence of multiple stable time crystal phases signifies a rich and diverse landscape for time crystal behavior \cite{sacha2017time,else2020discrete,kongkhambut2022observation,zaletel2023colloquium}. In addition, the RF field-driving technique plays an important role in creating more RF bands for the Rydberg state, thus providing a versatile method to obtain controllable many-body states.

Another interesting point is that the system transits from the time crystal phase to a chaotic phase during scanning of the voltage $U$, as depicted in area E in Fig. \ref{Distinct time crystal phases}(a) or panel (b4) in Fig. \ref{Distinct time crystal phases}(b). The irregular frequency distribution in panel (b4) of Fig. \ref{Distinct time crystal phases}(b) shows disordered properties that correspond to weak time translation symmetry breaking that results from imperfections in the orderly arrangement. This corresponds to the disordered time crystal, which is different to the regular time crystal. This is analogous to a conventional disordered crystal, i.e., a crystal with defects. This is of interest in materials science because these crystals can exhibit novel characteristics that are not observed in their perfect counterparts.

The distinct heights of the peaks in the Fourier spectrum [Fig. \ref{Distinct time crystal phases}(b2) and Fig.\ref{phase transition}(b)] indicate that more than one time crystal exists under the same physical conditions in the system. This shows that the system could represent the superposition of two time crystals. Because of the change in the interactions between the Rydberg atoms caused by $\Delta_c$, it becomes possible to manipulate the system's ground state to be the superposition of two time crystals with adjacent $f$. This finding may be helpful in study of the coherence and time-dependence of many-body states.
\\
\subsection*{Summary}

In summary, we have demonstrated the bifurcation of time crystals in strongly interacting Rydberg atoms driven using an external RF field. Through this RF field-driving approach, we were able to observe the intriguing symmetry breaking behavior of the time translation in Rydberg atoms. We observed multiple stable time crystals beyond the single time crystal within different coupling detuning ranges. By manipulating the RF field parameters, we observed a continuous phase transition from an LTO state to an STO state and discovered the bistability of the time crystal, which manifested as a hysteresis loop. The bifurcation of the time crystal in strongly interacting Rydberg atoms is attributed to the emergence of more complex temporal symmetries beyond the simple discrete time translation symmetry. Our work represents a milestone in understanding of the field of non-equilibrium transitions, and particularly with respect to the progression from a single time crystal phase to multiple distinct phases with different temporal symmetries.

\textit{Note.} While finishing this manuscript we became aware of a related work of reporting a phase transition from a continuous to a discrete time crystal [Ref.~\cite{kongkhambut2024observation}], and a related work (Ref.~\cite{jiao2024observation}) observing time crystal comb.

\bibliography{ref}

\begin{figure*}
\centering
\includegraphics[width=2.05\columnwidth]{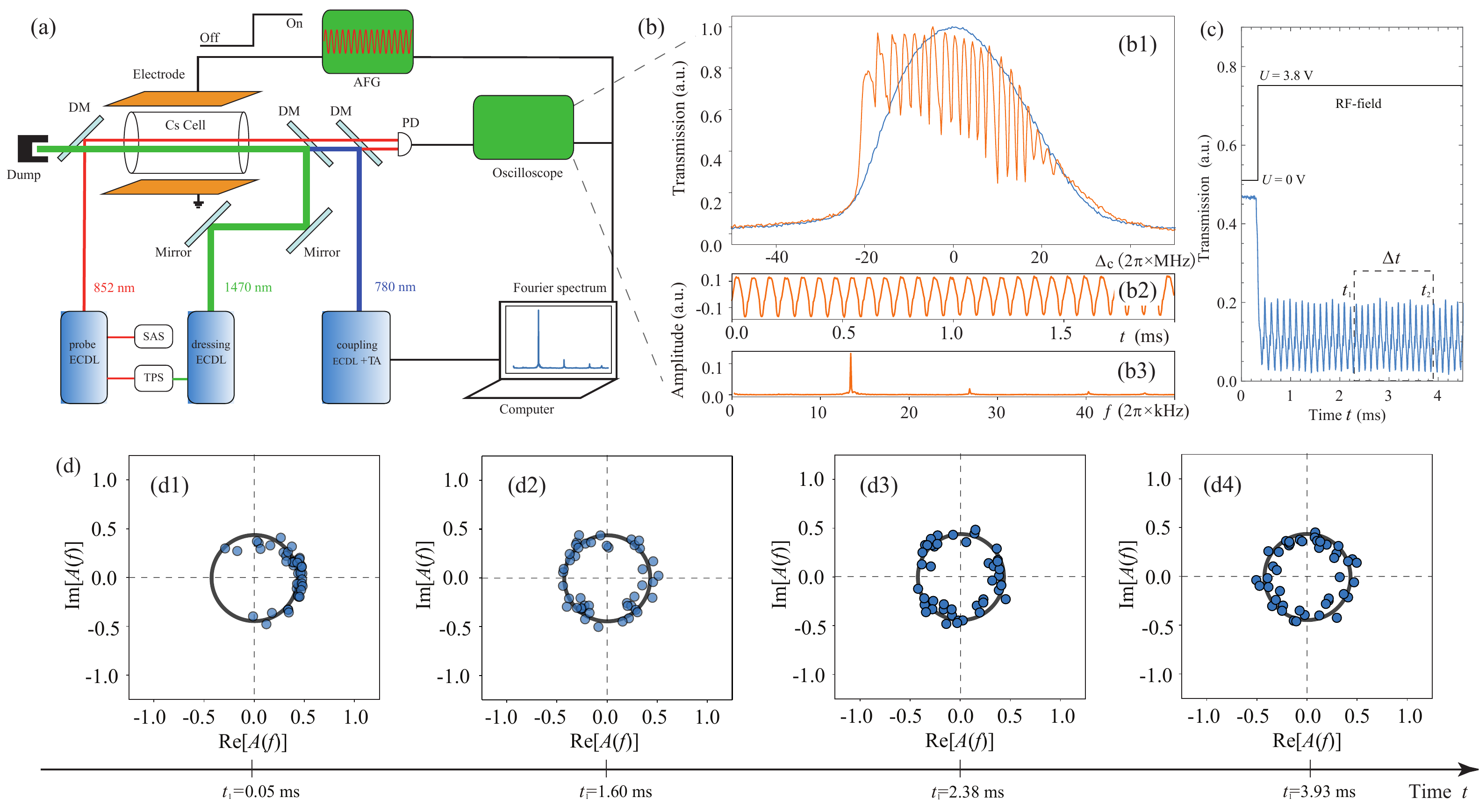}\\
\caption{\textbf{Details of experimental setup and probe transmission oscillations.} (a) Experimental setup. ECDL: external-cavity diode laser. PD: photoelectric detector. TA: tapered amplifier. AFG: arbitrary function generator. SAS: saturated absorption spectrum. DM: dichroic mirror. The dump is used to collect the optical beam. The computer is used to analyze the Fourier spectrum and visualize the data. (b) The EIT spectra with (orange) and without (blue) the external RF field are shown in (b1). Panel (b2) shows the oscillated EIT transmission in the time domain, and panel (b3) represents the Fourier spectrum of the time crystal. The first peak characterizes the frequency of the time crystal. (c) Quench dynamics of the probe transmission caused by switching of the RF field. (d) Measured distribution of the Fourier amplitudes with the dominant peak on the complex plane realized by selecting different starting times of $t_1 = 0.05$ ms (d1), 1.60 ms (d2), 2.38 ms (d3), and 3.93 ms (d4) within the time interval $\Delta t = 0.775$ ms.}
\label{fig.S1}
\end{figure*}

\section*{Methods}
\subsection*{Details of the experimental setup}
The experimental details are depicted in Fig.\ref{fig.S1}(a). In the experiments, we used the three-photon scheme to excite the Rydberg atoms. The 852 nm external-cavity diode laser (ECDL) was locked by using the saturated absorption spectrum (SAS) as a frequency reference signal. Another ECDL at 1470 nm was locked by using the two-photon spectrum (TPS) as a frequency reference signal. The 780 nm ECDL was amplified using a tapered amplifier (TA) as a coupling laser, where the power was 2 $W$. The probe laser is divided into two beams, where one beam serves as a reference and the other beam propagates in the opposite direction to the dressing laser and coupling laser beams. We used the arbitrary function generator (AFG) to generate the RF field and apply it to the electrodes. The probe transmission was recorded using the oscilloscope after the signal passed through the photoelectric detector (PD). To scan the external parameters and thus obtain the phase diagram of the system response, the coupling laser, the oscilloscope, and the AFG were all connected to a computer and controlled together. 

Panel (b1) in Fig.~\ref{fig.S1}(b) shows the measured EIT spectrum under the external RF field driving conditions, in which the frequency of the RF field was set at $\omega$ = $2\pi \times$8.1 MHz. In Fig.~\ref{fig.S1}(b), panels (b2) and (b3) represent the oscillated transmission and the Fourier transformation spectrum, respectively. The oscillated behavior was caused by the population of Rydberg atoms oscillating between two energy states with a specific period, creating a temporal order [with a frequency of $\sim2\pi \times$19 kHz], and displaying the characteristics of a continuous dissipative time crystal.

\begin{figure*}
\centering
\includegraphics[width=2\columnwidth]{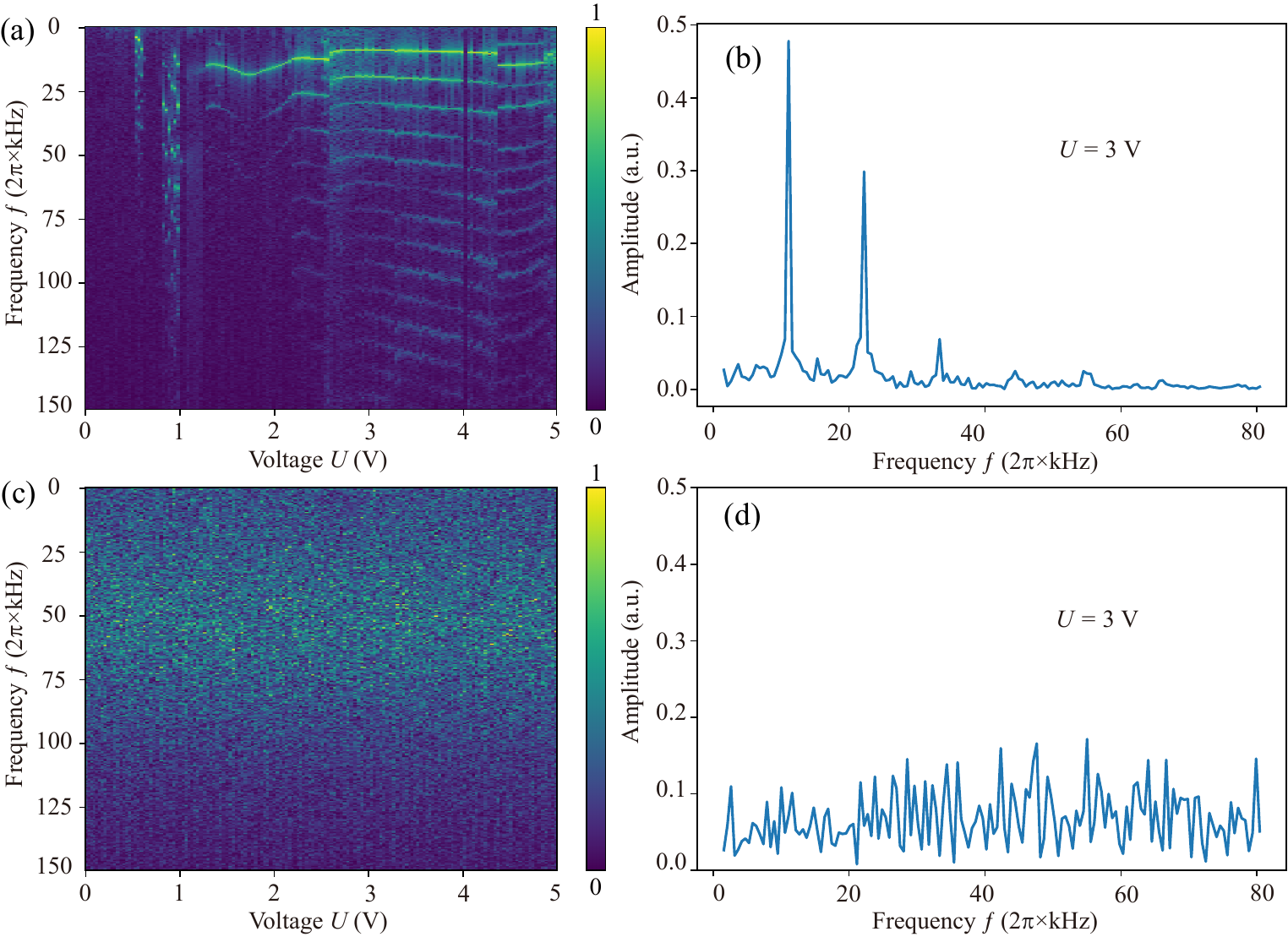}\\
\caption{\textbf{Normalized Fourier spectra with different probe intensities.} Fourier map of a section of the voltage $U$ range from 0 V to 5 V with a probe intensity of 64 $\mu W$ in (a) and of 3.5 $\mu W$ in (c). In (a), distinct peaks can be found, whereas only irregular noise is found in panel (c). The color bar ranging from 0 to 1 represents the Fourier transform intensity. (b) and (d) show the corresponding Fourier spectra at $U$ = 3 V.}
\label{fig.S2}
\end{figure*}

In the quench dynamics measurement [see Fig.~\ref{fig.S1}(c)], we measured the oscillated transmission by switching the RF field on. By selecting data points within the interval from $t_1$ $\sim$  $t_2$, as depicted in Fig.~\ref{fig.S1}(c), we obtained the real and imaginary parts of the amplitude of the dominant peak in the Fourier spectrum. In this process, we used 300 data points acquired from multiple experimental trials to construct the plots shown in Fig.\ref{setup} (e) and Fig.\ref{setup} (f).  

When the RF field is switched on, the system initially carries out an evolution that is not completely disordered but with the phase $\pi/2 \sim 3\pi/2$ missing, which means that the probe transmission oscillates within the phase distribution from $-\pi/2 \sim \pi/2$ at a small $t_1 = 0.05$ ms, as shown in Fig.~\ref{fig.S1}(d1). After an additional period of time, the system is completely out of order, as shown in Fig.~\ref{fig.S1}(d2)-(d4), and the oscillation then becomes completely random when $t_1$ = 3.93 ms. This occurs because the system is inclined to remain in a stationary phase at a small $t_1$, whereas it exhibits persistent oscillations with random phases at larger $t_1$ values, thus revealing the essence of spontaneous breaking of the time translation symmetry.

\subsection*{Comparison of different probe intensities}
The oscillated behavior during probe transmission indicates the existence of a time crystal, and this is closely related to the long-range interactions occurring between the Rydberg atoms. In the experiments, we observed evidence of this behavior by changing the population of the excited Rydberg atoms. Using the same measurement method that was used in the work in the main text, we obtained the Fourier spectrum of the system response, as shown in Fig.\ref{fig.S2}, by scanning the voltage from $U$ = 0 V to $U$ = 5 V at various probe field intensities. Therefore, we changed the probe intensities to excite different numbers of Rydberg atoms, and obtained different phase maps for the Fourier spectrum. When the probe intensity was set at 64 $\mu W$, the Fourier spectrum showed a series of peaks, as indicated in Fig.\ref{fig.S1}(a), which corresponded to distinct time crystal phases. In addition, we observed the time crystal phase of splitting and transitioning, which is caused by scanning of the voltage $U$ with an opposite direction, similar to that in Fig.\ref{phase transition} and Fig.\ref{Bistability} as described in the main text. Figure \ref{fig.S2} (b) shows the Fourier spectrum at $U$ = 3 V, where several peaks are present in the spectrum.

However, when the probe intensity decreases to 3.5 $\mu W$, the EIT spectrum still exists, but the oscillated behavior observed during probe transmission disappears. In addition, the Fourier spectrum of the system shows irregular noise, as illustrated in Fig.\ref{fig.S2} (c), without any peaks [see the example in Fig.\ref{fig.S2} (d) at $U$ = 3 V]. In this scenario, the number of Rydberg atoms is small because of the weak probing light, which results in weaker interactions between the Rydberg atoms. Furthermore, the oscillated behavior of the system disappears and the system then enters the normal phase from the time crystal phase. This suggests that the long-range interaction is responsible for the emergence of the time crystal.

\subsection*{Floquet theory of the coupling of Rydberg atoms and the RF field}
To describe the coupling between the RF field and the Rydberg atoms and to explain the reason for generation of the RF sidebands, we introduce the Floquet theory. In the Floquet treatment, the RF field is viewed as a periodic driving signal, and the Hamiltonian of the system is written as follows:
\begin{equation}
H(t) = H_0 + eU\rm{sin}(\omega t)
\end{equation}
where $H_0$ is the Hamiltonian of the atomic system in the absence of the field and $U$ is the voltage of the RF field. According to the Floquet theory \cite{Floquet}, the wave functions of a time-periodic Hamiltonian system can be written in the following form: 
\begin{equation}
\Psi_\nu(t)=e^{-i W_\nu t / \hbar} \psi_\nu(t)
\end{equation}
This wave function can be expanded using the standard basis states $|k\rangle$:
\begin{equation}
\Psi_\nu(t)=e^{-i W_\nu t / \hbar} \sum_k \sum_{m=-\infty}^{\infty} \tilde{C}_{\nu, k, m} e^{-i m \omega t}|k\rangle
\end{equation}
where $W_\nu$ are the Floquet quasi-energies, the integer $m$ represents the order of the sidebands, and $\tilde{C}_{\nu, k, m} $ is the Fourier expansion coefficient. The energies of the different sidebands are given by 
\begin{equation}
\hbar \omega_m=W_\nu+m \hbar \omega
\end{equation}
Under the RF electric field driving condition, the Rydberg energy levels form a series of equally spaced Floquet energy levels with the energy level spacing $\hbar \omega$. To model the oscillated spectrum of time crystal, we only consider the $\pm1$-order sidebands, as shown in Fig.\ref{setup}(a).

\subsection*{Mean-field theory description}
In our model, we consider $N$ interacting four-level atoms with a ground state $|g⟩$ and the Rydberg states $|R⟩$, $|R_1⟩$, and $|R_2⟩$ (all with an equal decay rate $\gamma$). Here, a mean-field treatment was introduced to simulate the results. The system's master equation is based on the multi Rydberg-state model \cite{wu2023observation}:
\begin{equation}
    \partial_t \hat{\rho} = i [\hat{H},\hat{\rho}] + \mathcal{L}_{R}[\hat{\rho}] + \mathcal{L}_{R_2}[\hat{\rho}] + \mathcal{L}_{R_2}[\hat{\rho}]
\end{equation}
where $\hat{H}$ is the Hamiltonian and $\mathcal{L}_{\{R,R_1,R_2\}}$ is the Lindblad jump operator. The Hamiltonian of the system is:
\begin{equation}
\begin{aligned}
    \hat{H} & =\frac{1}{2}\sum_{i}\left(\Omega\sigma_{i}^{gR}+\Omega_{1}\sigma_{i}^{gR_1}+\Omega_{2}\sigma_{i}^{gR_2}+h.c.\right)\\ & -\sum_{i}\left(\Delta n_{i}^{R}+\Delta_{1}n_{i}^{R_1}+\Delta_{2}n_{i}^{R_2}\right)\\ & +\sum_{i\neq j}\bigg[V_{ij}^{RR_1}n_{i}^{R}n_{j}^{R_1}+V_{ij}^{RR_2}n_{i}^{R}n_{j}^{R_2}+V_{ij}^{R_1R_2}n_{i}^{R_1}n_{j}^{R_2}\\ &+\frac{1}{2}(V_{ij}^{RR}n_{i}^{R}n_{j}^{R}+V_{ij}^{R_1R_1}n_{i}^{R_1}n_{j}^{R_1}+V_{ij}^{R_2R_2}n_{i}^{R_2}n_{j}^{R_2})\bigg]
\end{aligned}
\end{equation}
where $\sigma_{i}^{gr}$ ($r={R,R_1,R_2}$) represents the $i$-th atom transition between the ground state $\left| g \right\rangle$ and the Rydberg state $\left|  r \right\rangle$, $n_{i}^{R,R_1,R_2}$ are the population operators of the Rydberg energy levels $\left|  R \right\rangle$ and $\left|  R_1 \right\rangle$, and $\left|  R_2 \right\rangle$, and $V_{ij}^{RR_1}$, $V_{ij}^{RR_2}$, and $V_{ij}^{R_1R_2}$ are the interactions between the Rydberg atoms in states $\left|R\right\rangle$, $\left|R_{1}\right\rangle$, and $\left|R_{2}\right\rangle$, respectively. 

The Lindblad jump terms are given by:
\begin{equation}
    \mathcal{L}_r = (\gamma_{r}/2) \sum_i (\hat{\sigma}_i^{r g} \hat{\rho} \hat{\sigma}^{ gr}_i - \{\hat{n}_i^{r},\hat{\rho}\}),
\end{equation}
which represents the decay process from the Rydberg state $\left| r \right\rangle$  ($r={R,R_1,R_2}$) to the ground state $\left| g \right\rangle$. 

In the mean-field treatment, we obtained the following equations:
\begin{align}
    \dot{n}^R & =i \frac{\Omega}{2}\left(\sigma^{g R}-\sigma^{R g}\right)-\gamma n^R, \\
    \dot{n}^{{R}_1} & =i \frac{\Omega_1}{2}\left(\sigma^{g {R}_1}-\sigma^{{R}_1 g}\right)-\gamma n^{R_1}, \\
    \dot{n}^{{R}_2} & =i \frac{\Omega_2}{2}\left(\sigma^{g {R}_2}-\sigma^{{R}_2 g}\right)-\gamma n^{R_2}, \\
    \dot{\sigma}^{g R} & =i \frac{\Omega}{2}\left(2 n^R+ n^{R_1}+ n^{R_2}-1\right)+i\frac{\Omega_1}{2}\sigma^{R_1 R}\nonumber\\ &+i\frac{\Omega_2}{2}\sigma^{R_2 R} +i\left(\Delta-\Delta_{int}+i \frac{\gamma}{2}\right) \sigma^{g R}, \\
    \dot{\sigma}^{g {R}_1} & =i \frac{\Omega_1}{2}\left(2 n^{R_1}+ n^{R}+ n^{R_2}-1\right)+i\frac{\Omega}{2}\sigma^{R R_1}\nonumber\\
    & +i\frac{\Omega_2}{2}\sigma^{R_2 R_1}+i\left(\Delta_1-\Delta_{int}+i \frac{\gamma}{2}\right) \sigma^{g R_1}, \\
    \dot{\sigma}^{g {R}_2} & =i \frac{\Omega_2}{2}\left(2 n^{R_2}R+ n^{R_1}+ n^{R_12}-1\right)+i \frac{\Omega_1}{2}\sigma^{R_1 R_2} \nonumber\\
    & +i \frac{\Omega_1}{2}\sigma^{R_1 R_2}+i\left(\Delta_2-\Delta_{int}+i \frac{\gamma}{2}\right) \sigma^{g R_2}, \\
    \dot{\sigma}^{R R_1} & =i \left(\frac{\Omega_1}{2} \sigma^{g R_1}-\frac{\Omega_1}{2}\sigma^{R g}\right)-i\left(\Delta-\Delta_{1}-i \gamma\right) \sigma^{R R_1}, \\
     \dot{\sigma}^{R R_2} & =i \left(\frac{\Omega_2}{2} \sigma^{g R_2}-\frac{\Omega_2}{2}\sigma^{R g}\right)-i\left(\Delta-\Delta_{2}-i \gamma\right) \sigma^{R R_2},\\
    \dot{\sigma}^{R_1 R_2} & =i \left(\frac{\Omega_1}{2} \sigma^{g R_{2}}-\frac{\Omega_2}{2}\sigma^{R_{1}g}\right)-i\left(\Delta_{1}-\Delta_{2}-i \gamma\right) \sigma^{R_1 R_2}, 
\end{align}
where $\Delta_1 = \Delta - \omega$, $\Delta_2 = \Delta + \omega$, $\Delta_{int} = \xi(n^R+n^{R_1}+n^{R_2})$ is the interaction-induced shift, and $\xi$ is the interaction coefficient. In this way, we can calculate the system's dynamics within the limit cycle regime. We therefore obtained the population of the Rydberg states $n_{\rm{RR}}$, $n_{\rm{R_1R_1}}$, and $n_{\rm{R_2R_2}}$ versus time, as shown in Fig.\ref{fig.S1}. Figure.\ref{fig.S1}(a) shows the results obtained with no interaction, where the system will reach a steady state after a specified time [typically in the range from $\gamma t=0$ to $\gamma t=100$]. The Fourier spectrum of the data points from $\gamma t=100\sim200$ in Fig.\ref{fig.S1}(a) is shown in Fig.\ref{fig.S1}(b). Figure \ref{fig.S1}(c) shows the results obtained with interaction, in which the system achieves long-time oscillation, thus meaning that it enters a limit cycle phase. The Fourier spectrum of the data points from $\gamma t=100\sim200$ shown in Fig.\ref{fig.S1}(c) is given in Fig.\ref{fig.S1}(d). There are visible peaks in the amplitude [with different oscillation periodicities] in Fig.\ref{fig.S1}(d), thus revealing the distinct time crystals.

\begin{figure*}
\centering
\includegraphics[width=2\columnwidth]{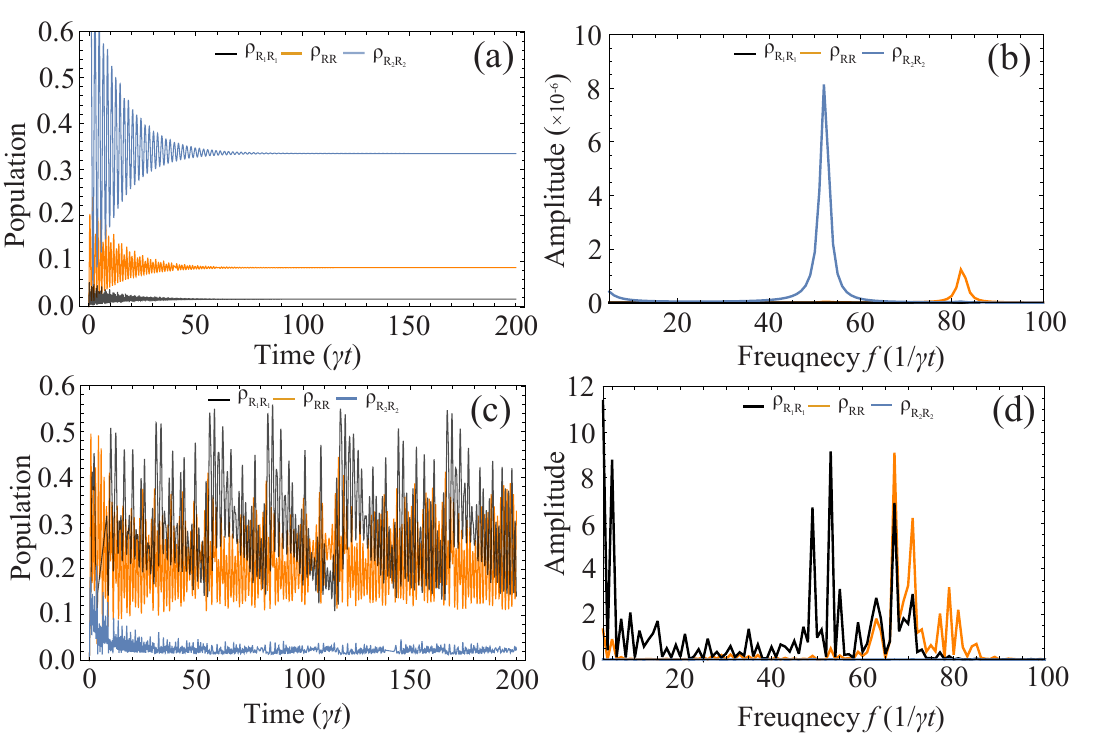}\\
\caption{\textbf{Theoretical simulation.} Calculated population of the Rydberg states $\rho_{\rm{RR}}$, $\rho_{\rm{R_1R_1}}$, and $\rho_{\rm{R_2R_2}}$ versus time with no interaction $\xi = 0$ (a) and with interaction $\xi = -16$ (c). The parameters $\Omega$ = $\Omega_1$ = $\Omega_2$ = 3, $\gamma = 0.1$, $w = 7$, and $\Delta = -5$ were selected here. (b) and (d) show the corresponding Fourier spectra of the data points from $\gamma t=100\sim200$ in (a) and (c), respectively. From these results, we find that the peak amplitude in (b) is much lower than that in (d).}
\label{fig.S3}
\end{figure*}

\section*{Acknowledgements}
D.-S.D. thanks for the previous discussions with Prof. Igor Lesanovsky and Prof. C. Stuart Adams on time crystals. We acknowledge funding from the National Key R and D Program of China (Grant No. 2022YFA1404002), the National Natural Science Foundation of China (Grant Nos. U20A20218, 61525504, and 61435011), the Anhui Initiative in Quantum Information Technologies (Grant No. AHY020200), and the Major Science and Technology Projects in Anhui Province (Grant No. 202203a13010001).

\section*{Author contributions statement}
D.-S.D. conceived the idea for the study. B.L. conducted the physical experiments and developed the theoretical model. The manuscript was written by D.-S.D and B.L. The research was supervised by D.-S.D. All authors contributed to discussions regarding the results and the analysis contained in the manuscript.

\section*{Competing interests}
The authors declare no competing interests.

\end{document}